\documentclass{article}
\begin{document}

\def\d{\partial}
\def\R{{\bf R}}
\def\f#1#2{\frac{#1}{#2}}
\def\be{\begin{equation}}
\def\ee{\end{equation}}
\def\bea{\begin{eqnarray}}
\def\eea{\end{eqnarray}}

\date{}

\title{Blow-up for solutions of hyperbolic PDE and spacetime singularities.}
\author{Alan D. Rendall \\
Max--Planck--Institut f\"ur Gravitationsphysik \\
Am M\"uhlenberg 1, 14476 Golm, Germany}

\maketitle

\begin{abstract} 
An important question in mathematical relativity theory is that
of the nature of spacetime singularities. The equations of general 
relativity, the Einstein equations, are essentially hyperbolic in nature
and the study of spacetime singularities is naturally related to blow-up
phenomena for nonlinear hyperbolic systems. These connections are
explained and recent progress in applying the theory of hyperbolic
equations in this field is presented. A direction which has turned 
out to be fruitful is that of constructing large families of solutions
of the Einstein equations with singularities of a simple type by solving 
singular hyperbolic systems. Heuristic considerations indicate, however,
that the generic case will be much more complicated and require
different techniques.
\end{abstract}   

\section{Introduction}

Anyone who has worked with nonlinear hyperbolic equations is familiar with
the phenomenon that smooth solutions often cease to exist after a finite
time. There are two common ways in which this happens. In the book of
Alinhac\cite{alinhac95} these have been called the \lq ODE blow-up 
mechanism\rq\ and the \lq geometric blow-up mechanism\rq. In the first case 
the solution itself is unbounded as the blow-up time is approached. In the 
second case the solution itself remains bounded while the first derivative 
is unbounded. Although the Einstein equations, which are the subject of
the following, are arguably the most geometric of hyperbolic differential
equations the blow-up which is typically encountered occurs by the ODE
mechanism in the terminology of \cite{alinhac95}. At the same time, the
question of the sense in which the Einstein equations can be called 
hyperbolic is delicate. These points are related, as will now be explained

The basic unknown in the Einstein equations is a pseudoriemannian metric
$g$. The equations are diffeomorphism invariant, in the sense that if a
metric $g$ on a manifold $M$ satisfies the Einstein equations and if
$\phi$ is a diffeomorphism of $M$ onto itself, then the pull-back
$\phi^*g$ is also a solution. In order to make contact with PDE theory
as it is usually formulated, it is necessary to choose a local 
coordinate system $x^\alpha$ on $M$ and treat the components 
$g_{\alpha\beta}$ of the 
metric in this coordinate system as the basic unknowns.
The Einstein equations imply a system of second order partial differential 
equations for these components. This system cannot have a well-posed Cauchy
problem because of diffeomorphism invariance. For suppose that 
$g_{\alpha\beta}$ is a solution of these coordinate equations with given
initial data. The functions $g_{\alpha\beta}$ are the components of a 
metric $g$. Now let $\phi$ be a diffeomorphism which leaves
a neighbourhood of the Cauchy surface invariant and let $\bar g=\phi^* g$.
Then $\bar g_{\alpha\beta}$ is a solution with the same initial data as 
$g_{\alpha\beta}$. However it is in general not equal to $g_{\alpha\beta}$. 
Thus uniqueness in the Cauchy problem in the usual sense fails for this system.
On the other hand, there is a different uniqueness statement which does hold.
It says that if $g$ and $\bar g$ are metrics with the same initial data on
a suitable Cauchy surface then there exists a diffeomorphism $\phi$ such 
that $\bar g=\phi^* g$ which is the identity on the Cauchy surface.
In the following the precise mathematical formulation of this statement 
will not be required, but the intuitive idea is important. It should be
noted that this uniqueness property, often called geometric uniqueness,
is exactly what is desired from the point of view of physics since
metrics related by a diffeomorphism should be regarded as physically
indistinguishable.

It has now been indicated that diffeomorphism invariance presents a
difficulty when studying the Cauchy problem for the Einstein equations.
How can this difficulty be overcome? In principle, one could attempt to
construct a diffeomorphism-invariant version of PDE theory but this has
apparently never been done. In practice, the procedure is to require 
some condition relating the coordinates to the metric. This breaks the
diffeomorphism invariance and opens up the possibility that with this
extra condition imposed the Einstein equations imply a system of
hyperbolic equations called the reduced equations, and that conversely
suitable
solutions of the reduced equations give rise to solutions of the Einstein
equations. Details on this procedure of hyperbolic reduction can be found
in \cite{friedrich00}. There are many different kinds of hyperbolic 
reduction and the task is to find reductions which are well-adapted 
to given problems.

The Einstein equations are the basic equations of general relativity,
which is the best existing theory of the gravitational field. Within
this theory spacetime singularities, such as the big bang, correspond to
singularities of $g_{\alpha\beta}$ which cannot be removed by a 
diffeomorphism, even one which itself becomes singular where 
$g_{\alpha\beta}$ does.

\section{The Gowdy equations}

A useful laboratory for studying questions concerning the global behaviour
of solutions of the Einstein equations is the class of Gowdy solutions.
These have high symmetry and after a suitable hyperbolic reduction give
the following system of semilinear wave equations:
\bea\label{gowdy}
-\d_t^2 X-t^{-1}\d_t X+\d_x^2 X&=&2(\d_t X\d_t Z-\d_x X\d_x Z)         \\
-\d_t^2 Z-t^{-1}\d_t Z+\d_x^2 Z&=&-e^{-2Z}((\d_t X)^2-(\d_x X)^2)\nonumber
\eea
Here $(t,x)\in (0,\infty)\times \R$ and $X$ and $Z$ are real-valued 
functions. Because of the symmetry only one space dimension remains.
It will be useful to compare this with the equation
\be\label{exp}
-\d_t^2 u+\d_x^2 u=-e^u
\ee
The equation (\ref{exp}) and its higher dimensional analogues have been 
studied in a series of papers by 
Kichenassamy and Littman\cite{kichenassamy93a, kichenassamy93b,
kichenassamy96a, kichenassamy96b, kichenassamy96c}. Partial results
of a similar type for (\ref{gowdy}) have been obtained in 
\cite{kichenassamy98a}
and \cite{rendall00a}. An obvious difference between the equations 
(\ref{gowdy}) and (\ref{exp}) is that while the singularities in solutions
of (\ref{exp}) develop dynamically in the $(t,x)$ plane there
is a singularity in the coefficients of (\ref{gowdy}). Moreover it was proved
by Moncrief\cite{moncrief80} that smooth Cauchy data for (\ref{gowdy}) 
at $t=t_0>0$
develop into a smooth solution on the whole of the interval $(0,\infty)$.
In fact this difference is more apparent than real. When proving theorems
about (\ref{exp}) it is convenient to introduce a function which is 
constant on the singular hypersurface and rewrite the equations in terms
of that. In reality, a corresponding procedure has been carried out when 
introducing the
coordinates used in the Gowdy equation. This is so natural in the context 
of the Einstein equations, where a priori all time coordinates are equally
valid, that it might pass unnoticed.

The nature of all these results is that they demonstrate the existence of
large families of solutions of the equations whose singularities can be
described precisely. This is achieved by using certain singular hyperbolic
equations, whose singularity is of the type known as Fuchsian.
The description of the singularity is given by an asymptotic expansion, 
which contains a number of free functions. These free
functions may be called \lq data on the singularity\rq. If these data are
analytic then the expansion actually converges. The case of analytic data
is easiest to handle. This was done for (\ref{exp}) in \cite{kichenassamy93a}
and \cite{kichenassamy93b} and for (\ref{gowdy}) in \cite{kichenassamy98a}.
The results for (\ref{exp}) were generalized to the case of smooth data
(in fact data with sufficiently high order Sobolev regularity) in
\cite{kichenassamy96a}. The theory developed there does not suffice to
cover (\ref{gowdy}) with data which are smooth but not analytic. This
difficulty was overcome in \cite{rendall00a}, where there is also a
general discussion of the issue of extending results using Fuchsian 
techniques from the analytic to the smooth case and a survey of the 
literature where these techniques have been applied to problems in
general relativity.

Now the asymptotic expansions for solutions of the Gowdy equations will
be presented for illustration.
\bea
X(t,x)&=&X_0(x)+t^{2k(x)}(\psi(x)+v(t,x))          \\
Z(t,x)&=&k(x)\log t+\phi(x)+u(t,x)
\eea
Here the functions $u(t,x)$ and $v(t,x)$ are remainder terms which are
$o(t)$ as $t\to 0$. Note the occurrence of powers of $t$ with exponents
depending on $x$ in the expansion for $X$. It is these which cause
additional technical difficulties in this case as compared to the case
of (\ref{exp}). Solutions of the above form are obtained for smooth data 
$X_0$, $k$, $\phi$ and $\psi$ which are arbitrary except for the 
restrictions that $0<k(x)<1$. The significance of the condition $k>0$ 
will not be discussed here. The condition $k<1$, known as the low 
velocity condition, will play a significant role in the following discussion.
It turns out that it can be removed at the expense of requiring
that $X_0$ should be constant. In this way a class of high velocity
solutions can be obtained. They depend, however, on one less free
function than the low velocity solutions.

For the physical interpretation of the results on singularities in 
Gowdy spacetimes it is important to know that the singularity in
the metric coefficients at $t=0$ cannot be removed by a diffeomorphism.
One way of doing this is to consider scalar polynomial quantities
in the curvature of the metric. These are real-valued functions
which are defined in a coordinate invariant way. One of the simplest
is the Kretschmann scalar $K$, which is the squared length (with respect
to the given metric) of the Riemann curvature tensor. It is a
function of $X$, $Z$ and their first and second order partial
derivatives whose explicit form is very complicated. Fortunately, for
a solution with an asymptotic expansion of the type given above it 
is not too difficult to calculate the leading term in the expansion of
$K$ about $t=0$. The result, discussed in \cite{kichenassamy98a}, is
\be
K\sim Q(k) t^{-3(k^2+1)/4}
\ee
for a certain quartic polynomial $Q$ whose only positive root is at
$k=1$. Thus, provided $k\ne 1$ everywhere, $K$ blows up uniformly as
$t\to 0$.

An important issue is that of the generality of the solutions constructed
by the above techniques. They are general in the crude sense that they 
depend on the same number of free functions as data for the same
equations on a regular Cauchy surface. In the case of (\ref{exp})
Kichenassamy proved in \cite{kichenassamy96b} using the Nash-Moser theorem 
that a non-empty open set of data on a regular Cauchy surface develop into
solutions of the type constructed by Fuchsian techniques. Up to now no
corresponding results have been proved in problems arising in general
relativity, such as the Gowdy problem. By a more direct approach, Chru\'sciel
\cite{chrusciel91} was able to show that a large open set of data 
for the Gowdy equations lead to solutions where the Kretschmann scalar blows 
up as $t\to 0$. Note that while the methods used in \cite{chrusciel91} were 
special to one space dimension, Fuchsian techniques do not depend 
essentially on the dimension.
 
\section{The Einstein-scalar field equations}

As mentioned above, the unknown of central importance in the Einstein
equations is a metric $g$. Physically, it contains the information about 
the gravitational field and hence about the mutual gravitational attraction
of material bodies. Thus it is natural that in any model involving the 
Einstein equations a description of the matter present is also necessary. In
fact, in general relativity the gravitational field has degrees of 
freedom which can propagate in the absence of matter. This leads to
the phenomenon of gravitational waves. The equations in the absence 
of matter are known as the vacuum Einstein equations and have the simple
geometric form ${\rm Ric}(g)=0$, where ${\rm Ric}(g)$ is the Ricci 
curvature of $g$. In the presence of matter these equations acquire a
non-vanishing right hand side. An example which will be important in the
following is that of the Einstein-scalar field equations. In that case
the Einstein equations take the form ${\rm Ric}(g)=\nabla\phi\otimes
\nabla\phi$. In general the matter fields must satisfy some equations
of motion and in the case of the scalar field the equation of motion
is that $\phi$ should satisfy the linear wave equation defined by the
metric $g$. In coordinate components this takes the form
\be
g^{\alpha\beta}(\d_\alpha\d_\beta\phi-\Gamma^\gamma_{\alpha\beta}\d_\gamma
\phi)=0
\ee
where 
\be
\Gamma^\gamma_{\alpha\beta}=\frac{1}{2}g^{\gamma\delta}
(\d_\alpha g_{\beta\delta}+\d_\beta g_{\alpha\delta}
-\d_\delta g_{\alpha\beta}) 
\ee
are the Christoffel symbols of $g$. Here we use the summation convention 
and $g^{\alpha\beta}$ is the inverse matrix of $g_{\alpha\beta}$.

It is known that in some cases the singularities of solutions of the 
Einstein equations display complicated oscillatory behaviour and it is
suspected that this occurs under very general circumstances. The best
known example of complicated behaviour is the Mixmaster solution. This
is a solution of the vacuum Einstein equations which is symmetric under
the group $SU(2)$. The Einstein equations reduce to ordinary differential
equations but the ODE solutions are very complicated. Until recently
our knowledge of the Mixmaster solution was based on numerical 
calculations and heuristic arguments. Now a number of features of
this picture have been proved rigorously by Ringstr\"om\cite{ringstrom00a}.
The generality of Mixmaster behaviour is still out of reach of present
analytical results. There have, however, been interesting advances in
numerical approaches. (See \cite{weaver98a}.) 

General predictions about the nature of singularities in solutions of 
the Einstein equations come from heuristic work of Belinskii, Khalatnikov
and Lifshitz\cite{bkl82}. According to their picture the singularities
in solutions of the vacuum Einstein equations should look like a
different Mixmaster solution at each spatial point near the singularity.
In particular the evolution at different spatial points decouples and
the solution of the PDE system can be approximated by solutions of an 
ODE system. (Here we are reminded of the ODE blow-up mechanism in the
terminology of Alinhac.) The same picture suggests that for the 
Einstein-scalar field system the decoupling should still take place
but the ODE solutions should now be monotone rather than oscillatory 
near the singularity. This gives rise to a hope that the Einstein-scalar 
field system should be easier to handle analytically than the Einstein 
vacuum equations. Following this lead, it was shown in \cite{andersson00}
that a Fuchsian analysis could be carried out for the Einstein-scalar
field system without any need for symmetry assumptions. Solutions are
obtained which depend on the same number of free functions as there
are in the Cauchy data on a regular Cauchy surface. In this crude sense
the solutions are general. Only the analytic case was treated up to now.

The asymptotic expansions obtained will now be presented.  This is only 
done for a certain diagonalizable case. This does not restrict the number
of free functions although it is a restriction on the solutions and a 
large part of the work of \cite{andersson00} was devoted to showing that 
the restriction is unnecessary. On the other hand, the diagonalizable
case is best suited to explain the results. For physical reasons we are
interested in solutions on a four-dimensional manifold. The time 
coordinate is denoted by $x^0$ and the spatial coordinates by $x^a$ with
$a=1,2,3$. The Gaussian coordinate conditions
\bea
g_{00}&=&-1                \\
g_{0a}&=&0                 
\eea
are imposed. The asymptotic expansions for the remaining unknowns are as 
follows:
\bea
g_{ab}&=&t^{2p_1}l_al_b+t^{2p_2}m_am_b+t^{2p_3}n_an_b+o(t^{q_{ab}}) \\
\phi&=&A\log t+B+o(1)     
\eea
Here $p_1$, $p_2$, $p_3$, $l_a$, $m_a$, $n_a$, $A$ and $B$ depend on the
spatial coordinates $x^a$ and $p_1+p_2+p_3=1$.
The $q_{ab}$ are suitably chosen real numbers,
which will not be discussed in detail here. The important thing is that 
they represent higher order terms. The functions occurring in the expansion
must satisfy certain algebraic and differential equations which will also 
not be discussed, except to say that there is a good understanding of the
solution set of these equations. There is an equivalent of the condition
$k<1$ in the Gowdy case, namely that all $p_a$ should be positive. This
implies that all $p_a$ are smaller than one. 

The conclusion of this work is that it provides strong support for the
ideas of Belinskii, Khalatnikov and Lifshitz in the case that a scalar field
is present. The conceptual, geometric and calculational aspects of the proof
are a lot more complicated than in the Gowdy case, but the analytic input
is almost identical.

\section{The nonlinear scalar field}

From the foregoing discussion of the great difference in the nature of
singularities in solutions of the Einstein equations caused by the presence 
of a scalar field the impression might arise that the whole question 
depends in a delicate way on the exact nature of the matter fields
present. In fact this impression is probably false: there should probably
only be a small number of universality classes which cover most types of
matter. This will be illustrated by a class of matter models which will
be referred to collectively as the nonlinear scalar field. It turns 
out that a wide variety of choices of the parameters which go into the
definition all lead to the same asymptotic behaviour, which is that 
of the (linear) scalar field discussed in the previous section. This will
also allow some of the aspects of the Fuchsian theory to be shown in 
action.

The main interest for physicists of the equations obtained by coupling the 
Einstein equations to a nonlinear scalar field has been the the
phenomenon known as inflation. The mathematics of relevance to inflation
concerns the behaviour of the solutions in the direction away from the
singularity and so has no direct connection with the discussion below.

Consider first the following equation
\be\label{flatwave}
-\d_t^2\phi-t^{-1}\d_t\phi+\Delta\phi=m^2\phi+V'(\phi)
\ee
Here $\phi$ is a real-valued function, $\Delta$ is the Laplacian on 
$\R^n$ (the case of $n=3$ is of most interest for the following) and $m$ 
is a constant (mass) which we may as well take to be non-negative. The 
function $V$ (potential) is smooth and $V'$ vanishes at least 
quadratically
at the origin. The separation of the right hand side into the sum of a 
mass term and a potential term has no essential significance in this 
problem. It is convenient for making contact with the notation used in 
the physics literature. The aim is to determine whether this equation 
has solutions which near $t=0$ have the leading order asymptotics 
$A(x)\log t+B(x)$ already encountered in the previous section. To this
end, introduce $\psi=t^{-\epsilon}(\phi-A\log t-B)$, where $\epsilon$ is
a positive constant, whose purpose will be seen later on. When 
reexpressed in terms of $\psi$ the equation (\ref{flatwave}) becomes:
\bea\label{flatwave2}
-t^2\d_t^2\psi-(2\epsilon+1)t\d_t\psi
&-\epsilon^2\psi+t^2\Delta\psi=-t^{2-\epsilon}\log t\Delta A
-t^{2-\epsilon}\Delta B
\nonumber\\
&+t^{2-\epsilon}m^2(A\log t+B+t^{\epsilon}\psi)+W(t,x,\psi)
\eea
where $W(t,x,\psi)=t^{2-\epsilon}V'(A\log t+B+t^\epsilon\psi)$. The 
equation has been multiplied
by $t^2$ so as to make the combination $td/dt$ appear. Some choices of 
$V$ to be found in the physics literature are $V_1(\phi)=V_0$, a constant,
$V_2(\phi)=\phi^\alpha$, $V_3(\phi)=e^{\beta\phi}$ for positive constants 
$\alpha$ 
and $\beta$. The value of the constant $V_0$ has no effect on the equation
(\ref{flatwave}) but does have an effect when the coupling to the Einstein
equations is considered. The potential $V_1$ combined with a non-zero
value of $m$ occurs in a model known as chaotic inflation. The potential
$V_3$ occurs in what is called power-law inflation. (It is not the potential
itself which has power-law behaviour, but the long-time behaviour of the 
solution.) The functions $W$ corresponding to $V_1$, $V_2$ and $V_3$ 
are $W_1(t,x,\psi)=0$, 
$W_2(t,x,\psi)=\alpha t^{2-\epsilon}(A\log t+B+t^\epsilon\psi)^{\alpha-1}$ 
and $W_3(t,x,\psi)=\beta e^{\beta B} t^{2-\epsilon+\beta A} 
\exp(\beta t^\epsilon\psi)$. 
The functions $W_1$, $W_2$ and $W_3$ are smooth for $t>0$. Suppose that the 
function $A$ is nowhere vanishing. Then $W_1$ and $W_2$ extend continuously 
to $t=0$ by defining them to be zero there. This is also true of their 
derivatives with respect to the arguments $x$ and $\psi$. This will be 
abbreviated by saying that the functions $W_1$ and $W_2$ are regular. As 
for $W_3$, can be made regular in this sense by choosing $\epsilon$ 
sufficiently small provided $A>-2\beta^{-1}$. This is a restriction on
the data on the singularity. In particular, if $\beta$ is negative then 
it bounds $A$ away from zero. If $V$ had been chosen to grow even faster
at infinity, e.g. $V(\phi)=e^{\phi^2}$ then the corresponding $W$ could 
not have been made regular by imposing inequalities on $A$. This type of
potential with faster then exponential growth does not seem to occur in
the physics literature and the Fuchsian theory does not seem to apply to
it. It will not be considered further here. 

The Fuchsian theory we will use concerns first order systems and thus 
the first step in the analysis is to reduce (\ref{flatwave2}) to first
order in a suitable way. Let $v=t\d_t\psi+\epsilon\psi$ and $w=t\d_x\psi$. 
Then the the following system is obtained:
\bea\label{flatwave3}
t\d_t\psi+\epsilon\psi-v&=&0\nonumber                                  \\
t\d_t w+(\epsilon-1)w&=&t\d_x v                              \\
t\d_t v+\epsilon v&=&t\d_x w+\Delta A t^{2-\epsilon}\log t
+t^{2-\epsilon}\Delta B   \nonumber\\
&&-t^{2-\epsilon}m^2(A\log t+B+\psi)+W(t,x,\psi)\nonumber
\eea
This system is of the following general form for Fuchsian equations:
\be
t\d_t u+N(x)u=t^\eta f(t,x,u,\d_x u)
\ee
for some $\eta>0$.
It is also symmetric hyperbolic, a luxury which is usually not easy to 
obtain in problems in this area. On the other hand, it does have a
disadvantage. The theorem proved in \cite{kichenassamy98a} required the 
assumption that $t^{N(x)}$ be bounded for $t$ near zero. To obtain this
the eigenvalues of $N(x)$ should have non-negative real parts. 
Since $\epsilon$ has to be chosen small in general, depending on
the data, the eigenvalue $\epsilon-1$ will be negative and this condition
is not satisfied for the equation (\ref{flatwave3}). An alternative form 
is obtained by substituting the definition of $v$ back into 
(\ref{flatwave3}) in one place. The result is:
\bea\label{flatwave4}
t\d_t\psi+\epsilon\psi-v&=&0\nonumber                            \\
t\d_t w+\epsilon w&=&t\d_x v+t\d_x \psi                              \\
t\d_t v+\epsilon v&=&t\d_x w+\Delta A t^{2-\epsilon}\log t
+t^{2-\epsilon}\Delta B   \nonumber\\
&&-t^{2-\epsilon}m^2(A\log t+B+\psi)+W(t,x,\psi)\nonumber
\eea
The equation (\ref{flatwave4}) is not symmetric hyperbolic. It does,
however satisfy the conditions of the existence theorem for the case
of analytic data proved in \cite{kichenassamy98a}. In order to ensure
this it is important that $\epsilon$ is positive. If $\epsilon$ were
chosen to be zero then all the eigenvalues of the matrix $N(x)$ would
vanish but the matrix would not be diagonalizable. This would violate 
one of the conditions assumed in \cite{kichenassamy98a}, namely that
zero eigenvalues, if they occur at all, only occur through $N(x)$ 
being a direct sum of a matrix whose eigenvalues have positive real 
parts with a zero matrix.

The result of the previous discussion is that given analytic functions $A$
and $B$ with $A$ nowhere vanishing then if the restrictions on $V$ and $A$
already obtained are satisfied there exists a unique solution, analytic
in $x$ and continuous in $t$ of (\ref{flatwave}) with the desired 
asymptotics. This shows the existence of a family of solutions 
depending on two free analytic functions, which is the same as the
number of functions which can be given as data on a regular Cauchy 
surface. 

In \cite{rendall00a} a method for going beyond the analytic case in
Fuchsian problems was applied to the Gowdy equations. Since the
matrix $N(x)$ occurring in (\ref{flatwave4}) does not depend on $x$
it should be possible to prove an existence theorem for that equation
with smooth data given on the singularity by the methods of
\cite{kichenassamy96a}. Here we would like to note that the method
of \cite{rendall00a} generalizes easily to the case of smooth data for
(\ref{flatwave}). The main idea is to use both the systems (\ref{flatwave3})
and (\ref{flatwave4}). If the smooth data is approximated by a sequence
of analytic data then a corresponding sequence of analytic solutions
can be produced using (\ref{flatwave3}). These define a sequence of 
analytic solutions of (\ref{flatwave4}). Then the fact that the latter
equation is symmetric hyperbolic allows the use of energy estimates
to prove that the sequence of analytic solutions converges to the 
desired smooth solution. Details of the procedure can be found in
\cite{rendall00a}. 

\section{Coupling the nonlinear scalar field to the Einstein equations}

In this section it will be shown that the Einstein equations coupled to
the nonlinear scalar field can be handled by the same method as that used
for the case with $m=0$ and $V=0$ in \cite{andersson00}. The arguments will
only be sketched since most of the steps are identical. The potential 
will be assumed to be one of the functions $V_1$, $V_2$ or $V_3$ 
considered in the last section. In addition it is assumed that $A\ne 0$
and that in the case of $V_3$ the inequality $A>-2\beta^{-1}$ holds.
The Einstein equations take the form
\be
{\rm Ric}(g)=\nabla\phi\otimes\nabla\phi+[\f12 m^2\phi^2+V(\phi)]g
\ee
Evidently this equation does change if $V$ is changed by an additive
constant. The equation for the scalar field is
\be\label{wave}
g^{\alpha\beta}(\d_\alpha\d_\beta\phi-\Gamma^\gamma_{\alpha\beta}\d_\gamma
\phi)=m^2\phi+V'(\phi)
\ee
Applying the Gaussian coordinate conditions $g_{00}=-1$ and $g_{0a}=0$
gives
\be
g^{\alpha\beta}\d_\alpha\d_\beta\phi=-\d_t^2\phi+g^{ab}\d_a\d_b\phi
\ee
and the close relation of equation (\ref{wave}) to equation (\ref{flatwave})
begins to become apparent. We are looking for solutions to the Einstein
equations coupled to a nonlinear scalar field where the metric has the
asymptotic form found in \cite{andersson00} for the case of a linear 
scalar field. This form implies that the quantities $t^2 g^{ab}$ all
vanish like some positive power of $t$ as $t\to 0$ and thus do not 
disturb the Fuchsian form of the wave equation. The same holds for the
quantities $t\Gamma^a$ and the quantity $t\Gamma^0$ is equal to
one up to higher order terms. These facts cannot be easily seen from
the equations given in this paper. To check them it is necessary to
use more detailed results from \cite{andersson00}. The conclusion is
that up to higher order terms the equation (\ref{wave}) agrees with the
model equation (\ref{flatwave}).

What remains to be done in order to check that the Einstein equations 
coupled to a nonlinear scalar field have solutions which behave 
asymptotically like those for a linear scalar field is to see that
certain source terms in the Einstein equations are equal to those in
the case with a linear massless scalar field (which has already been 
solved in \cite{andersson00}), up to terms of higher order in $t$. The 
essential thing is to estimate the quantity $\d_t^2\phi+\f12 
m^2\phi^2+V(\phi)$. More specifically it is necessary to estimate the
deviation of this quantity from the corresponding quantity with $\psi=0$
and show that it is $O(t^{-2+\alpha_0})$ where $\alpha_0$ is a positive
parameter introduced in \cite{andersson00}. In the course of the proofs 
in that paper it is permissible to reduce the size of the number 
$\alpha_0$ if required. It is easy to check that, under the assumptions 
already made on $V$ and $A$, the parameter $\alpha_0$ can be chosen so
as to ensure this estimate. This is the same kind of procedure as
choosing $\epsilon$ small in the case of the nonlinear scalar field.
The information obtained in this way must be used in two places. It
must be used in the Einstein constraints in estimating the quantity
$\bar C$ of \cite{andersson00}. It must also be used in evolution
equations for $\kappa^a{}_b$ in \cite{andersson00}, which constitute
the essential
part of the Einstein evolution equations. The additional contribution 
which arises when passing from the massless linear scalar field to the
nonlinear scalar field does not change the equations for $a\ne b$. For
$a=b$ the coefficient $\alpha_0$ appears. It can be concluded from all
this that solutions of the Einstein equations 
coupled to a nonlinear scalar field which have the same asymptotics as 
in the case of a massless linear scalar field and which are of the same
degree of generality can be constructed by Fuchsian techniques. This has
been shown under the assumption that the potential is of one of the forms
$V_1$, $V_2$ or $V_3$ and that in the third case the initial datum $A$
satisfies an additional restriction. In this proof the velocity-dominated
system, which is the system of equations satisfied by the leading order
approximation to the solution, is the same as in the case of the massless
linear scalar field.

\section{Formation of localized structure}

It has already been indicated that the use of Fuchsian theory to
construct spacetimes with singularities of a well-understood type
is not likely to be sufficient to handle general singularities of
solutions of the Einstein equations, due to the appearance of
oscillatory  behaviour in time as the singularity is approached.
There is, however, another effect which complicates the situation 
in an essential way. Note that the solutions constructed by Fuchsian
techniques are such that the solution blows up like $\log t$ as $t=0$
is approached and the spatial derivatives of the solutions of all
orders are also $O(\log t)$. Taking spatial derivatives does not
increase the rate of blow-up. It will now be shown that there are 
solutions for which this property does not hold. Consider the 
transformation:
\bea
\tilde X&=&X/(X^2+e^{2Z})  \\
\tilde Z&=&\log [e^Z/(X^2+e^{2Z})]
\eea
This transformation is equal to its own inverse and transforms solutions
of the Gowdy equations into solutions. The idea is now to take a 
solution $(\tilde X,\tilde Z)$ with the low velocity asymptotic
behaviour and transform it to a new solution $(X,Z)$.

The case will be considered that $\tilde X_0 (0)=0$, $\tilde X_0 (x)\ne 0$
for $x\ne 0$ and $\d_x \tilde X_0 (0)\ne 0$. For the transformed solution 
$X^2+e^{2Z}$ is of the form 
\be
(\tilde X_0)^2+(e^{2\tilde\phi}+2\tilde X_0\tilde\psi)t^{2\tilde k}
+o(t^{2\tilde k}) 
\ee
From this it is not hard to see that $X$ is bounded on an interval
about $x=0$ uniformly in $t$. Similarly $Z=O(\log t)$. Thus the 
transformed solution $(X,Z)$ does not blow up any faster than the 
original solution $(\tilde X,\tilde Z)$. However spatial derivatives
blow up faster in general. To see this it suffices to evaluate
$Z(t,0)$ and $Z(t,t^{\tilde k/2})$. The second of these is bounded for values
of $x$ close to $x=0$ as $t\to 0$ while the first behaves like 
$-\tilde k\log t$. It follows from the mean value theorem that the 
maximum value of $\d_x Z$ in a small region about $x=0$ must blow up 
at least as fast as $t^{-\tilde k/2}$. Of course precise rates of blow-up
could be calculated for this and higher order derivatives if desired.
It is seen that these solutions of the Gowdy equations, unlike those 
constructed by the Fuchsian method, have spatial derivatives which blow 
up faster than the function itself. There is formation of localized 
structure.

In the explicit example given above the localized structures formed
do not have an invariant geometrical meaning and are due to a bad choice
of variables. On the other hand, there is convincing numerical evidence
due to Berger, Moncrief and collaborators for a second kind of localized
structure. In the latter case there is no reason to doubt that there is 
a real geometrical effect involved. For a survey of this work and
some pictures of the structures, see \cite{berger97}. There is also a
lot of interesting work on the subject using heuristic methods. A
rigorous treatment is still lacking. 

Although this is still in the domain of speculation, there are
arguments that in more general situations things will get yet more 
complicated. The basic idea will now be sketched. The oscillatory
behaviour in the Mixmaster model is composed of a series of bounces
where expansion turns into contraction and vice versa. In the
inhomogeneous case bounces of this kind are supposed to happen 
independently at different spatial points. However this does not
always happen coherently. There are spatial regions of coherent 
behaviour but these split into smaller regions as the singularity
is approached. On the boundaries between regions spatial derivatives
become large. The localized features observed in the Gowdy solutions
are simple examples of boundaries of this kind. The heuristic
models available suggest that in the Gowdy case this kind of 
formation of structure will only happen a finite number of times, at
least for generic solutions. In the general case, on the other hand,
it is expected to happen infinitely many times before the singularity
is reached, creating structures on arbitrarily small spatial scales.
This behaviour has been called \lq spacetime turbulence\rq. Information
on this can be found in \cite{montani95}. It has also been suggested 
that this process which is predicted by the picture of Belinskii,
Khalatnikov and Lifshitz actually shows that the picture must 
eventually break down in the generic case before the singularity is
reached. 

It seems clear that it will take a long time before the ideas mentioned
in the last paragraph can be brought into the fold of well-understood
mathematical phenomena. Nevertheless, the recent progress described in
this paper is a ground for optimism that there will be significant 
mathematical developments in the area of singularities of solutions
of the Einstein equations in the next few years.

\end{document}